\documentclass[aps,prl,twocolumn,groupedaddress,showpacs,floatfix]{revtex4-1}
\usepackage{graphics}
\usepackage{epsfig}
\usepackage{graphicx}
\usepackage{dcolumn}
\begin{document}
\title{First Direct Measurement of the $^{17}$O(p,$\gamma$)$^{18}$F Reaction Cross-Section at Gamow Energies for Classical Novae}


\author{
	D.A.\,Scott,$^1$ 
	A.\,Caciolli,$^{2,3}$ 
	A.\,Di Leva,$^{4}$
	A.\,Formicola,$^{5*}$
	M.\,Aliotta,$^1$	
	M.\,Anders,$^6$	
	D.\,Bemmerer,$^6$	
	C.\,Broggini,$^2$	
	M.\,Campeggio,$^7$	
	P.\,Corvisiero,$^8$	
	Z.\,Elekes,	$^6$	
	Zs.\,F\"ul\"op,$^9$	
	G.\,Gervino,$^{10}$
	A.\,Guglielmetti,$^7$
	C.\,Gustavino,$^5$
	Gy.\,Gy\"urky,$^9$
	G.\,Imbriani,$^4$
	M.\,Junker,$^5$
	M.\,Laubenstein,$^5$
	R.\,Menegazzo,$^2$
	M.\,Marta,$^{11}$
	E.\,Napolitani,$^{12}$
	P.\,Prati,$^8$
	V.\,Rigato,$^3$
	V.\,Roca,$^4$
	E.\,Somorjai,$^9$
	C.\,Salvo,$^{5,8}$
	O.\,Straniero,$^{14}$
	F.\,Strieder,$^{13}$
	T.\,Sz\"ucs,$^9$
	F.\,Terrasi,$^{15}$
	D.\,Trezzi$^{16}$\\
	(LUNA Collaboration)
	}

\affiliation{$^1$SUPA, School of Physics and Astronomy, University of Edinburgh, Edinburgh EH9 3JZ, UK} 
\affiliation{	$^2$INFN, Sezione di Padova, 35131 Padova, Italy} 
\affiliation{$^3$INFN, Laboratori Nazionali di Legnaro, Padova, Italy} 
\affiliation{	$^4$Dipartimento di Scienze Fisiche, Universit\`a di Napoli ``Federico II'', and INFN, Sezione di Napoli, Napoli, Italy} 
\affiliation{	$^5$INFN, Laboratori Nazionali del Gran Sasso, Assergi, Italy} 
\affiliation{	$^6$Helmholtz-Zentrum Dresden-Rossendorf, Dresden, Germany} 
\affiliation{	$^7$Universit\`a degli Studi di Milano and INFN, Sezione di Milano, Milano, Italy} 
\affiliation{	$^8$Dipartimento di Fisica, Universit\`a di Genova, and INFN, Genova, Italy} 
\affiliation{	$^9$ATOMKI, Debrecen, Hungary} 
\affiliation{	$^{10}$Dipartimento di Fisica Sperimentale, Universit\`a degli Studi di Torino, and INFN, Sezione di Torino, Torino, Italy} 
\affiliation{$^{11}$GSI Helmholtzzentrum f\"ur Schwerionenforschung GmbH, 64291 Darmstadt, Germany}
\affiliation{	$^{12}$MATIS-IMM-CNR at Dipartimento di Fisica e Astronomia, Universit\`a di Padova, Padova, Italy}%
\affiliation{	$^{13}$Institut f\"ur Experimentalphysik III, Ruhr-Universit\"at Bochum, Bochum, Germany} 
\affiliation{	$^{14}$Osservatorio Astronomico di Collurania, Teramo, and INFN, Sezione di Napoli, Napoli, Italy} 
\affiliation{	$^{15}$Seconda Universit\`a di Napoli, Caserta, and INFN, Sezione di Napoli, Napoli, Italy} 
\affiliation{	$^{16}$INFN, Sezione di Milano, Milano, Italy} 
	
\date{\today}

\begin{abstract}
Classical novae are important contributors to the abundances of key isotopes, such as the radioactive $^{18}$F,  whose observation by satellite missions could provide constraints on nucleosynthesis models in novae.
The $^{17}$O(p,$\gamma$)$^{18}$F reaction plays a critical role in the synthesis of both oxygen and fluorine isotopes but its reaction rate is not well determined because of the lack of experimental data at energies relevant to novae explosions. In this study, the reaction cross section has been measured directly for the first time in a wide energy range $E_{\rm cm} \simeq 200-370~{\rm keV}$ appropriate to hydrogen burning in classical novae. In addition, the $E$=183~keV resonance strength, $\omega \gamma$=1.67$\pm$0.12~$\mu$eV, has been measured with the highest precision to date. The uncertainty on the $^{17}$O(p,$\gamma$)$^{18}$F reaction rate has been reduced by a factor of 4, thus leading to firmer constraints on accurate models of novae nucleosynthesis.
\end{abstract}

\pacs{
{26.20.Cd;}
{26.30.-k;}
{26.50.+x}
}

\maketitle

Classical novae, a frequent phenomenon in our Galaxy, are explained as thermonuclear explosions on the surface of  white dwarf stars accreting hydrogen-rich material from less evolved companions in binary star systems \cite{pac65} and have been proposed as a key source of $^{13}$C, $^{15}$N, $^{17,18}$O, and $^{18,19}$F isotopes in the Universe \cite{jos98}. In particular, the short-lived radioisotope $^{18}$F ($t_{1/2}=110~{\rm min}$) may provide a signature of novae outbursts through the detection of 511 keV $\gamma$-ray emission from positron-electron annihilation following its $\beta^+$ decay. Indeed, the observation of these $\gamma$ rays by satellite missions could put constraints on current nova models \cite{her99}. 
Hydrogen burning of $^{17}$O is believed to play a key role on the destruction of $^{17}$O and on the formation of $^{18}$F, mainly through the competing reactions $^{17}$O(p,$\gamma$)$^{18}$F and $^{17}$O(p,$\alpha$)$^{14}$N. 
Thus, the thermonuclear rates of both reactions should be determined with a high degree of accuracy directly in the energy region of hydrogen burning in classical novae.
In this Letter, we report on a four-fold improvement in the $^{17}$O(p,$\gamma$)$^{18}$F reaction rate determination. We have measured the $^{17}$O(p,$\gamma$)$^{18}$F reaction cross section down to the lowest energies to date and within the Gamow window \cite{ili07} for peak temperatures $T = 0.1-0.4~{\rm GK}$ \cite{jos98} .

At nova temperatures, the $^{17}$O(p,$\gamma$)$^{18}$F reaction rate is dominated by a direct-capture (DC) reaction mechanism despite the presence of two narrow resonances at $E=66$ and $183~{\rm keV}$ above the proton threshold in $^{18}$F \cite{fox05,new10} (all energies are in the center-of-mass system, unless otherwise stated). In addition, non-resonant contributions arise from the low-energy tails of two broad resonances at $E= 556.7$ and $676.7~{\rm keV}$. 
A reliable determination of the total $^{17}$O(p,$\gamma$)$^{18}$F reaction rate thus requires the accurate knowledge of the individual energy dependence of both resonant and non-resonant contributions.
\begin{figure*}[ht]
\includegraphics[scale=0.9]{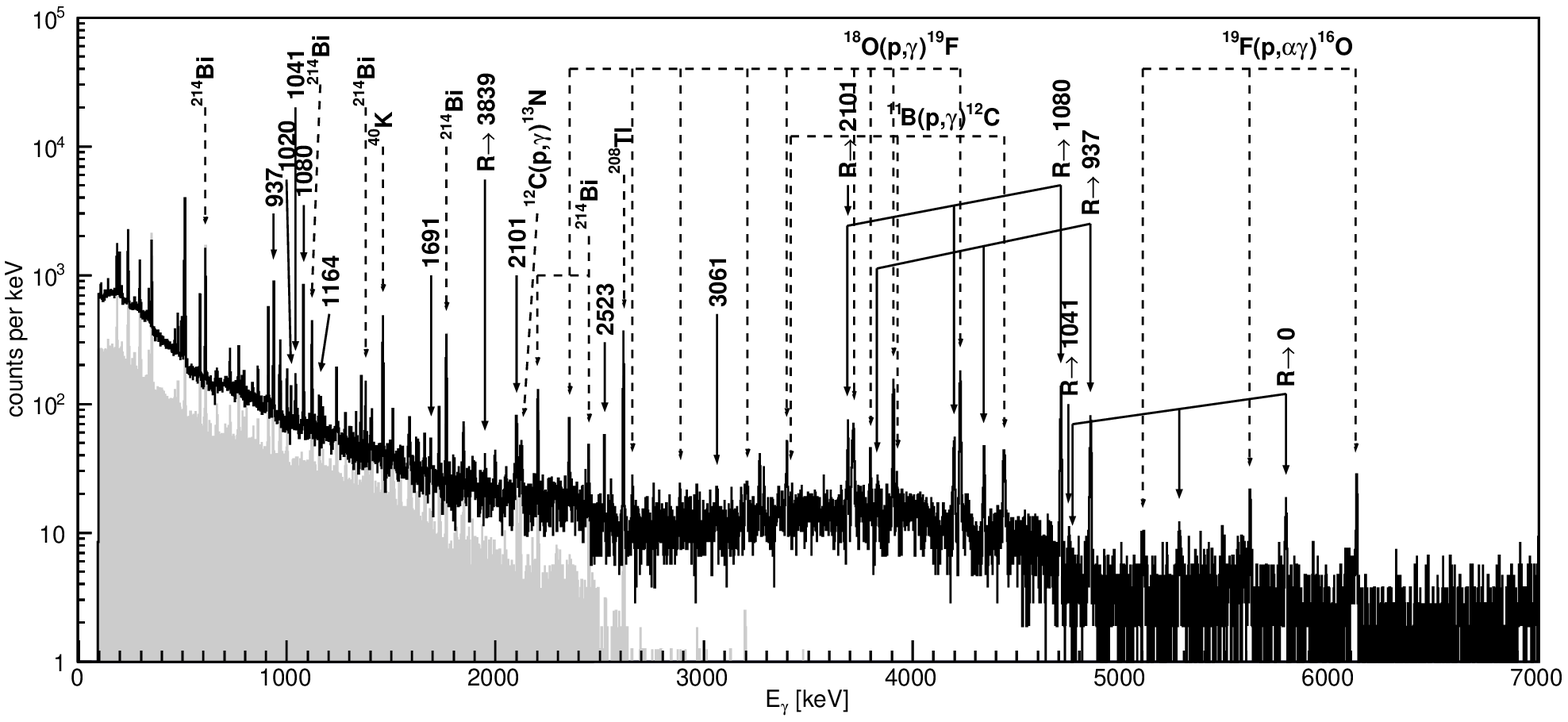}
\flushleft \caption{Sample spectrum from prompt $\gamma$-ray measurements of the  $^{17}$O(p,$\gamma$)$^{18}$F reaction at the $E=183~{\rm keV}$ resonance. Primary transitions are labelled as R$\rightarrow$$E_{\rm x}$, while secondary transitions are labelled by their energies. Some of the most prominent background transitions (dashed arrows) are also shown for comparison, together with a time-normalised background spectrum (grey area).}
\label{gammaspec}
\end{figure*}

Early investigations by Rolfs \cite{rol73} reported a constant DC contribution to the S-factor (S$_{\rm DC}$) \cite{ili07,sfac} in agreement with the four lowest data points measured at energies $E=280-425~{\rm keV}$ by prompt $\gamma$-ray detection \cite{rol73}. It was later questioned by Fox {\em et al.} \cite{fox05} whether these data points were dominated by the DC process or affected by the presence of the two broad resonance tails. Thus, in Ref. \cite{fox05}, the S$_{\rm DC}$ was calculated using measured spectroscopic factors and a realistic Woods-Saxon potential and found to be up to a factor of $\sim$2.5 lower than that reported in Ref. \cite{rol73}.
A subsequent measurement \cite{new10} at $E \simeq 257-470~{\rm keV}$ led to an S$_{\rm DC}$ in agreement with the predictions by Fox {\em et al.} \cite{fox05} and still about a factor of 2 lower than in Ref. \cite{rol73}. 
More recently, the total (i.e., DC plus broad-resonance contributions) S factor was measured at 
$E = 260-470~{\rm keV}$ using the DRAGON recoil separator at TRIUMF \cite{hag12} and found to be in fairly good agreement  with values by Ref. \cite{new10}, but consistently higher than the total S factor reported in Ref. \cite{fox05}.
As for the $E=183~{\rm keV}$ resonance strength, only two values exist to date: $\omega \gamma = (1.2\pm0.2)\times10^{-6}~{\rm eV}$, as determined by a prompt $\gamma$-ray measurement \cite{fox04,fox05}, and $\omega \gamma = (2.2\pm0.4)\times10^{-6}~{\rm eV}$, as determined by the activation technique \cite{cha05}, disagree at the 95\% confidence level. The origin of this discrepancy is not understood at present, but may be due in part to unobserved gamma transitions in Refs. \cite{fox04,fox05} and/or an inappropriate subtraction of the DC component in either  Refs. \cite{fox04,fox05} or  Refs. \cite{cha05}. 
Thus, the lack of experimental data at low energies and the largely unconstrained S$_{\rm DC}$-factor have so far precluded the accurate determination of the thermonuclear rate for this important reaction.

Here, we report on the results of a new and improved investigation of the $^{17}$O(p,$\gamma$)$^{18}$F reaction using both activation and prompt $\gamma$-ray detection techniques to address a key source of discrepancy in existing results. Measurements were carried out at the 400kV LUNA accelerator \cite{for03} of the Laboratory for Underground Nuclear Astrophysics (LUNA) facility, which offers significant improvements in sensitivity thanks to its low-background environment \cite{cos09}. 
A proton beam, with currents up to 200~$\mu$A on target, entered the target chamber through a liquid-nitrogen cooled copper-pipe biased to -300 V for secondary electrons suppression. The target was directly water cooled with de-ionised water. Targets were prepared by anodization of Ta backings (0.3 mm thick disks) in isotopically enriched water (66\% in $^{17}$O and 4\% in $^{18}$O). Full details on target preparation and characterisation have been reported in \cite{cac12}. The target thickness was closely monitored for signs of degradation during intense proton-beam bombardment by regularly measuring the thick-target yield profile \cite{ili07} of the narrow isolated resonance at $E=143~{\rm keV}$ in $^{18}$O(p,$\gamma$)$^{19}$F \cite{cac12}. 
Prompt $\gamma$ rays from the $^{17}$O(p,$\gamma$)$^{18}$F reaction ($Q$=5606.5$\pm$0.5$~{\rm keV}$) were detected using a large volume (115\% relative efficiency) high-purity germanium (HPGe) detector placed at 1.5~cm from the target and surrounded by 5 cm lead. Both detector and target were at an angle of 55$^{\circ}$ with respect to the beam axis, with the detector's front face parallel to the target surface. 
Energy calibration, full-energy peak efficiency, and total efficiency were determined taking into account corrections for true-coincidence summing as described in Ref. \cite{lim10}. Further details on the data analysis not reported here will be presented in a forthcoming publication \cite{next-LUNA-paper}.

\begin{figure}[t]
\includegraphics[scale=0.35]{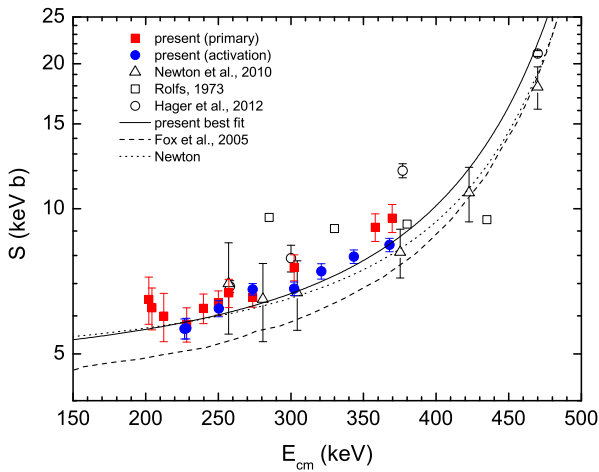}
\caption{Total astrophysical S factor as a function of center-of-mass energy for the $^{17}$O(p,$\gamma$)$^{18}$F reaction. Filled symbols refer to the activation (circles) and the prompt $\gamma$-ray (squares) measurements in the present study. The solid line is the best fit to our data, while the dotted line labeled ``Newton" is calculated using the DC contribution from Ref. \cite{new10} and the resonance parameters from Ref. \cite{fox05} as used in Ref. \cite{ili10b} (see text for further details).}
\label{fig:S-exp}
\end{figure}

The $^{17}$O(p,$\gamma$)$^{18}$F reaction proceeds by populating several states in $^{18}$F, leading to a complex decay scheme. Measurements were taken at the $183~{\rm keV}$ resonance and at several off-resonance energies.  At LUNA it was possible to observe and identify a large number of the $\gamma$-ray transitions from the entrance channel to $^{18}$F intermediate or ground state (the {\em primaries}) and subsequent direct or multistep decay of excited states to the $^{18}$F ground state (the {\em secondaries}). A sample $\gamma$-ray spectrum showing the quality of our data is given in Fig. \ref{gammaspec}. Independent determinations of the total S-factor were carried out for all primary and secondary transitions. 
The main primary transitions to $^{18}$F  states at $E_{\rm x}=937, 1121, 3061, 3791, 3839$ and $4116\,{\rm keV}$ were analysed using the peak-shape approach described in Ref. \cite{imb05}. 
The resulting total astrophysical S-factor, obtained from the sum of all primary contributions, is shown in Fig. \ref{fig:S-exp}. 
In total, 14 secondaries were also observed and analysed leading to an S factor (not shown in the Figure) in agreement with that from the primaries. Summing corrections amounted on average to about 30\% and were properly taken into account.
In addition to the statistical errors shown, the primary total S-factor data carry a systematic uncertainty due to summing effects (1.5\%), relative (1.5\%) and absolute (3.5\%) efficiency uncertainties, and an additional contribution of 7.6\% (in common with the activation data) obtained as a sum in quadrature of uncertainties on collected charge (3\%), stopping power (4\%), stoichiometry (3\%), target thickness (3.8\%), and $^{17}$O abundance (3\%). 

The $E$=183~keV resonance strength $\omega \gamma$ was determined using the thick-target yield approach \cite{ili07}. Prior to our investigation the $E$=183~keV resonance had been observed to decay to only two states at $E_{\rm x}$=1080 and 937 keV in $^{18}$F, with branching ratios of 60\% and 40\%, respectively \cite{til95}. However, several additional transitions were identified in our study, as shown in Fig. \ref{gammaspec}. 
Both the resonance strength and the branching ratios of all observed transitions were treated as free parameters in a global $\chi^2$ fit to the experimental counts as in Ref. \cite{imb05}. 
For the three most dominant transitions, R$\rightarrow$1080, R$\rightarrow$937 and R$\rightarrow$2101 keV, the branching ratios were determined as 41$\pm$3\%, 25$\pm2$\% and 12$\pm2$\%, respectively. The branching ratios for a remaining six observed transitions were less than 6\% each. Non-resonant contributions were subtracted for each transition individually. The measured $E$=183~keV resonance strength was $\omega\gamma$=1.70$\pm$3\%(stat)$\pm$7.7\%(sys)$\mu$eV. (Note that for measurements at the resonance, the target thickness uncertainty is only relevant for the DC subtraction and the common uncertainties reduce to 6.6\%). 

An independent determination of both resonant and non-resonant contributions to the $^{17}$O(p,$\gamma$)$^{18}$F total cross section was also obtained with the activation technique, i.e. by detecting the 511 keV $\gamma$ rays from the positron annihilation following the $\beta^+$ decay of $^{18}$F.
Targets were irradiated for several hours to saturate the $^{18}$F activity, using the same setup described earlier but without lead shielding.
Loss of $^{18}$F from the target during irradiation was estimated to be below 1\% at all bombarding energies according to GEANT4 Monte Carlo simulations.
Activation spectra were recorded every 20 minutes over a period of several hours using the low-background facility STELLA (SubTErranean Low Level Assay) \cite{arp96}. The absolute efficiency of the detector was measured with a calibrated $^{85}$Sr source which emits a single $\gamma$-ray line at 514 keV allowing for a direct determination of the efficiency at the required energy, free from summing corrections. 
The laboratory background was negligible compared to the count rate of the activated targets. 
The irradiated targets were placed on top of the detector crystal and a Ta absorber, in direct contact with the activated target, was used to fully stop the emitted positrons.
Our best value of  $t_{1/2}=109.6 \pm 0.4~$min for the $^{18}$F half-life was obtained as an average of several fits to experimental data at different energies and found to be in excellent agreement with the literature value $t_{1/2} =109.77 \pm 0.05~$min. 
The only contaminant to the spectra was found to be the short-lived $^{13}$N $\beta^+$ emitter.
No longer-lived positron emitters were observed after several half-lives. From the number of observed disintegrations at off-resonance energies, the total S factor was determined following a standard procedure described in Ref. \cite{ili07}.  Results from the activation analysis are shown in Fig. \ref{fig:S-exp}. In addition to the statistical errors shown, data are affected by an independent systematic uncertainty of 2.9\% (sum in quadrature of uncertainties from backscattering losses (1\%), $^{13}$N contamination (1\%), detector efficiency (2.5\%)), and by a 7.6\% uncertainty in common with the prompt $\gamma$-ray data (see above).
The $E$=183 keV resonance strength was derived from measurements at energies on the plateau of the thick-target yield after subtraction of a DC contribution of about 18\%, which results in a 3\% DC-correction uncertainty on $\omega \gamma$.
The obtained resonance strength, $\omega\gamma$=1.65$\pm$1.8\%(stat)$\pm$7.8\%(sys) $\mu$eV, is in excellent agreement with the result of the prompt $\gamma$-ray measurements. 
The weighted average of the resonance strength values from both measurements gives $\omega \gamma$=1.67$\pm$0.12(stat+sys) $\mu$eV. 

\begin{table}
  \caption{\label{tab:rate} Reaction rate for the $^{17}$O(p,$\gamma$)$^{18}$F$_{\rm gs}$ reaction as a function of temperature.}
  \begin{ruledtabular}
    \begin{tabular}{ c c c c }
\footnotesize
Temperature & lower limit & recommended value & upper limit \\
T (GK) & \multicolumn{3}{c}{N$_{A} <\sigma\upsilon>$ (cm$^{3}$ mol$^{-1}$ s$^{-1}$)} \\

\hline

0.040	& $	2.15\times10^{-12	}  $ & $	2.50\times10^{-12	}  $ & $	2.90\times10^{-12	}  $ \\
0.050	& $	6.62\times10^{-11	}  $ & $	7.71\times10^{-11	}  $ & $	8.96\times10^{-11	}  $ \\
0.060	& $	6.66\times10^{-10	}  $ & $	7.69\times10^{-10	}  $ & $	8.87\times10^{-10	}  $ \\
0.070	& $	3.63\times10^{-09	}  $ & $	4.13\times10^{-09	}  $ & $	4.71\times10^{-09	}  $ \\
0.080	& $	1.40\times10^{-08	}  $ & $	1.57\times10^{-08	}  $ & $	1.76\times10^{-08	}  $ \\
0.090	& $	4.47\times10^{-08	}  $ & $	4.91\times10^{-08	}  $ & $	5.40\times10^{-08	}  $ \\
0.100	& $	1.27\times10^{-07	}  $ & $	1.38\times10^{-07	}  $ & $	1.50\times10^{-07	}  $ \\
0.110	& $	3.40\times10^{-07	}  $ & $	3.67\times10^{-07	}  $ & $	3.96\times10^{-07	}  $ \\
0.120	& $	8.57\times10^{-07	}  $ & $	9.22\times10^{-07	}  $ & $	9.92\times10^{-07	}  $ \\
0.130	& $	2.05\times10^{-06	}  $ & $	2.20\times10^{-06	}  $ & $	2.36\times10^{-06	}  $ \\
0.140	& $	4.60\times10^{-06	}  $ & $	4.92\times10^{-06	}  $ & $	5.26\times10^{-06	}  $ \\
0.150	& $	9.67\times10^{-06	}  $ & $	1.03\times10^{-05	}  $ & $	1.10\times10^{-05	}  $ \\
0.160	& $	1.91\times10^{-05	}  $ & $	2.03\times10^{-05	}  $ & $	2.17\times10^{-05	}  $ \\
0.180	& $	6.22\times10^{-05	}  $ & $	6.62\times10^{-05	}  $ & $	7.05\times10^{-05	}  $ \\
0.200	& $	1.67\times10^{-04	}  $ & $	1.78\times10^{-04	}  $ & $	1.89\times10^{-04	}  $ \\
0.250	& $	1.10\times10^{-03	}  $ & $	1.17\times10^{-03	}  $ & $	1.25\times10^{-03	}  $ \\
0.300	& $	4.70\times10^{-03	}  $ & $	5.04\times10^{-03	}  $ & $	5.41\times10^{-03	}  $ \\
0.350	& $	1.88\times10^{-02	}  $ & $	2.04\times10^{-02	}  $ & $	2.21\times10^{-02	}  $ \\
0.400	& $	7.65\times10^{-02	}  $ & $	8.44\times10^{-02	}  $ & $	9.29\times10^{-02	}  $ \\
0.450	& $	2.83\times10^{-01	}  $ & $	3.16\times10^{-01	}  $ & $	3.51\times10^{-01	}  $ \\
0.500	& $	8.80\times10^{-01	}  $ & $	9.88\times10^{-01	}  $ & $	1.10\times10^{+00}  $ \\
0.600	& $	5.18\times10^{+00}  $ & $	5.82\times10^{+00}  $ & $	6.51\times10^{+00}  $ \\
0.700	& $	1.87\times10^{+01}  $ & $	2.09\times10^{+01}  $ & $	2.34\times10^{+01}  $ \\
0.800	& $	4.87\times10^{+01}  $ & $	5.44\times10^{+01}  $ & $	6.06\times10^{+01}  $ \\
0.900	& $	1.02\times10^{+02}  $ & $	1.13\times10^{+02}  $ & $	1.26\times10^{+02}  $ \\
1.000	& $	1.82\times10^{+02}  $ & $	2.02\times10^{+02}  $ & $	2.24\times10^{+02}  $ \\
    \end{tabular}
  \end{ruledtabular}
\end{table}

To obtain an overall total S factor, the activation and the prompt $\gamma$-ray data sets were analyzed in a common fit procedure. Unfortunately, inclusion of other data sets (e.g. from Ref \cite{hag12}) in our fit was not possible since these seem to be affected by a mixture of systematic and point-to-point uncertainties that prevent a common and self-consistent analysis of all data sets.
Following the phenomenological approach of Refs. \cite{fox05,ili10b}, the fitting function included contributions from the two broad resonances at $E$=557 and 667~keV, with energy dependence of the involved partial widths described by the broad-resonance formalism  \cite{ili07}, as well as a constant DC term. 
The weak energy dependence of the DC component was neglected in the present analysis, as in previous works \cite{fox05,ili10b}, since its effect is beyond the precision of our data. Finally, preliminary calculations \cite{next-LUNA-paper} indicate that interference effects between resonant and DC contributions are negligible and have not been included.
A proper treatment of the systematic uncertainties \cite{dag94} required the introduction of three scaling factors, one for each data set to account for non-common systematic uncertainties, and one for the common uncertainties. Their values were then obtained simultaneously in a modified $\chi^2$ fit (for a similar approach see Ref. \cite{sch12}). 
Additional free parameters in the fit were the DC contribution and the partial $\Gamma_\gamma$ width of the 557 keV resonance. The former parameter was left unconstrained while the latter was treated as a free parameter weighed by a Gaussian probability distribution with expectation value and standard deviation given by the literature value 
$\Gamma_{\gamma,557}=0.57\pm0.13$~eV \cite{ili10c}.
The influence of the second broad-resonance on the fit quality was extremely small and, thus, $\Gamma_{\gamma,677}$ was fixed to the literature value \cite{ili10c}. 
The best fit ($\chi^2=13.9$ for 19 data points and 5 fit parameters) was obtained for $\Gamma_{\gamma,557}=0.70$~eV and a DC component of $\rm S_{DC}=4.4\pm0.4$~keVb with scaling factors of $c_{\rm act}=1.008$, $c_{\rm prim}=0.972$, and $c_{\rm com}=0.963$. These factors are in good agreement with their expectation value of 1, within the corresponding systematic uncertainties they represent.
\begin{figure}[t*]
\includegraphics[scale=0.4]{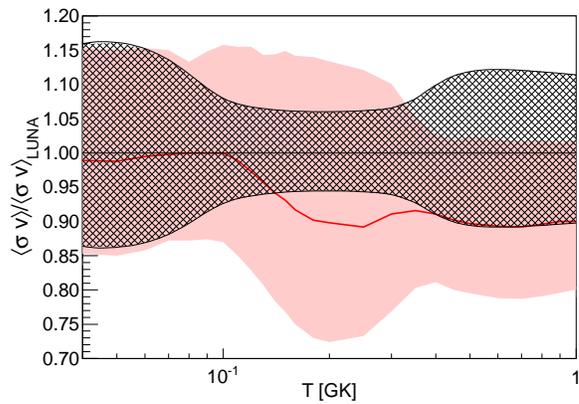}
\caption{(Colour online) $^{17}$O(p,$\gamma$)$^{18}$F reaction rate ratio as a function of temperature. The (red) solid line is the ratio between the rate from a recent compilation rate (Ref. \cite{ili10b}) and the present value. Hatched and shaded areas represent a 1$\sigma$ uncertainty on present and previous rate, respectively. In the relevant nova temperatures $T=0.1-0.4~{\rm GK}$, the rate uncertainty is reduced by a factor of 4 in the present study.}
\label{fig:rate}
\end{figure}

The astrophysical $^{17}$O(p,$\gamma$)$^{18}$F reaction rate was calculated using the formalism described in Ref. \cite{ili10a} as an incoherent sum of contributions from two narrow resonances (the $E=183~{\rm keV}$ one from the present work, the $E=66~{\rm keV}$ one from the literature \cite{ili10c}) and the combined contribution of the broad resonances and the direct capture, this latter obtained as numerical integration of the present S-factor best fit (solid curve in Fig. \ref{fig:S-exp}). The resulting reaction rate is tabulated in Table \ref{tab:rate} with lower and upper limits given by the 68\% confidence level. A comparison between our rate and that from a recent compilation \cite{ili10b} (Fig. \ref{fig:rate}) shows an improvement of a factor of 4 in the present rate uncertainty.

On the basis of our improved rate, we have explored preliminary implications on the abundances of key isotopes produced by classical novae.  
For example, according to Ref. \cite{ili02} a 25\% (1$\sigma$) uncertainty in the $^{17}$O(p,$\gamma$)$^{18}$F reaction rate leads to a $40-50$\% variation on the calculated yields of $^{18}$O, $^{18}$F and $^{19}$F at $T=0.1-0.4~{\rm GK}$.
With the present rate being determined at the 5\% level (see Fig. \ref{fig:rate}), the uncertainty on the $^{18}$O, $^{18}$F and $^{19}$F abundances reduces to less than 10\% and puts firmer constraints towards more accurate nucleosynthesis calculations in novae events. The effects of our revised rate on the computation of detailed nova models will be discussed in a forthcoming paper.

In summary, the $^{17}$O(p,$\gamma$)$^{18}$F reaction cross section has been measured for the first time in the  relevant energy region of hydrogen burning in classical novae. The astrophysical S factor has been determined both by means of prompt $\gamma$-ray and activation measurements and results from the two approaches were found to be in excellent agreement. In addition, we have obtained the most accurate determination of the $E$=183 keV resonance strength to date as $\omega\gamma$=1.67$\pm$0.12 $\mu$eV. The reaction rate uncertainty was reduced by a factor of 4 leading to improved constraints on classical novae nucleosynthesis.

The authors wish to thank the INFN mechanical workshop, electronics and chemical laboratories of LNGS, as well and Dr H. Costantini and Dr C. Mazzocchi for their contributions to the early stages of the experiments. AC acknowledges financial support by Fondazione Cassa di Risparmio di Padova e Rovigo. ADL acknowledges financial support by MIUR (FIRB RBFR08549F). Financial support by OTKA K101328, NN83261, DFG (BE 4100/2-1), and NAVI is also gratefully acknowledged.

$^*$ Corresponding author: alba.formicola@lngs.infn.it


\begin{thebibliography}{30}
\bibitem{sfac} The astrophysical S factor is defined as ${\rm S}(E)=\sigma(E)E{\rm exp(2\pi\eta)}$, with  $\sigma$ the cross section, $\eta$ the Sommerfeld parameter, and $E$ the interaction energy \cite{ili07}.
\end{thebibliography}
\end{document}